\documentclass[aps]{revtex4}
\usepackage{epsfig,amsmath,amsfonts,wasysym}
\usepackage{dsfont}
\usepackage{graphicx}
\newcommand{\fr}[2]{{\frac{#1}{ #2}}}
\newcommand{\Ref}[1]{(\ref{#1})}
\newcommand{\be}{\begin{equation}}
\newcommand{\ee}{\end{equation}}
\newcommand{\bn}{\begin{eqnarray}}
\newcommand{\en}{\end{eqnarray}}
\newcommand{\bd}{\begin{displaymath} }
\newcommand{\ed}{\end{displaymath}}
\newcommand{\bnn}{\begin{eqnarray*}}
\newcommand{\enn}{\end{eqnarray*}}
\newcommand{\bs}{\begin{subequations}}
\newcommand{\es}{\end{subequations}}
\newcommand{\adb}{\allowdisplaybreaks }

\begin{document}

\title{Self-force of a point charge in the space-time of a symmetric wormhole}

\author{Nail R. Khusnutdinov\footnote{e-mail: nrk@kazan-spu.ru}${}^{a,b}$
and \ Ilya V. Bakhmatov\footnote{e-mail: ilya.bakhmatov@gmail.com}${}^{a}$
}

\address{${}^a$Department of Physics, Kazan State University, Kremlevskaya 18,
Kazan 420008, Russia \\ ${}^b$Department of Physics, Tatar State
University of Humanity and Pedagogic, Tatarstan 2, Kazan 420021,
Russia}


\begin{abstract}
We consider the self-energy and the self-force for an electrically
charged particle at rest in the wormhole space-time. We develop
general approach and apply it to two specific profiles of the
wormhole throat with singular and with smooth curvature. The
self-force for these two profiles is found in manifest form; it is
an attractive force. We also find an expression for the self-force
in the case of arbitrary symmetric throat profile. Far from the
throat the self-force is always attractive.
\end{abstract}

\maketitle

\section{Introduction}

Wormholes are topological handles linking different regions of the
Universe or different universes. The activity in wormhole physics
was first initiated by the classical paper by Einstein and Rosen
\cite{EinRos35}, and later by Wheeler \cite{WheBook}. The latest
growth of interest in wormholes was connected with the ``time
machine''\/, introduced by Morris, Thorne and Yurtsever in Refs.
\cite{MorTho88,MorThoYur88}. Their works led to a surge of activity
in wormhole physics \cite{VisBook}. The main and unsolved problem in
wormhole physics is whether wormholes exist or not. The wormhole has
to violate energy conditions and the source of the wormhole geometry
should be exotic matter. One example \cite{Kra00,KhuSus02,Gar05} of
such exotic matter is quantum fluctuations, which may violate the
energy conditions. Another possible sources are a scalar field with
reversed sign of kinetic term \cite{Arm02}, and cosmic phantom
energy \cite{Sus05,Lob05}. The question of wormhole's stability is
not simple and requires subtle calculations (see, for example
\cite{Lob05_2}). The problem arises because a wormhole needs some
amount of exotic matter which violates energy conditions and has
unusual properties. Recently it was observed that some observational
features of black holes can be closely mimicked by spherically
symmetric static wormholes \cite{DamSol07} having no event horizon.
Some astronomical observations indicate possible existence of black
holes (see, for example, Ref. \cite{Nar05}). It is therefore
important to consider possible astronomical evidences of the
wormholes. There is an interesting observation of quasar Q0957+561
which shows the existence of a compact object without an event
horizon. Some aspects of wormholes' astrophysics were considered in
Ref. \cite{KarNovSha06}, where it was noted that this massive
compact object may correspond to a wormhole of macroscopic size with
strong magnetic field. A matter may go in and come back out of the
wormhole's throat.

In the framework of general relativity there exists a specific
interaction of particles with gravitating objects -- the
gravitationally induced self-interaction force which may have
considerable effect on the wormholes' physics. It is well-known
\cite{DeWBre60} that in a curved background alongside with the
standard Abraham-Lorenz-Dirac self-force there exists a specific
force acting on a charged particle. This force is the
manifestation of non-local essence of the electromagnetic field.
It was considered in details in some specific space-times (see
Refs. \cite{Poi,Khu05} for review). For example, in the case of
the straight cosmic string space-time the self-force appears to be
the only form of interaction between the particle and the string.
Cosmic string has no Newtonian potential but nevertheless a
massless charged particle is repelled by the string \cite{Lin86},
whereas massive uncharged particle is attracted by the string due
to the self-force \cite{Gal90}. The non-trivial internal structure
of the string does not change this conclusion \cite{KhuBez01}. The
potential barrier appears which prevents the charged particle from
penetrating into the string. For GUT cosmic strings the potential
barrier is $\sim 10^5 Gev$. The wormhole is an example of the
space-time with non-trivial topology. The consideration of the
Casimir effect for a sphere that surrounds the wormhole's throat
demonstrates an unusual behavior of the Casimir force
\cite{KhaKhuSus06} -- it may change its sign depending on the
radius of the sphere. It is expected that the self-force in the
wormhole space-time will show an unusual behaviour too. As we will
show in next sections the self-force indeed has an unusual
behaviour -- the charged particle is attracted by the wormhole
instead of expected repulsive action. This result is valid for
arbitrary profile of the throat. Therefore charged particles are
likely to gather at the wormhole's throat.

The organization of this paper is as follows. In Sec. \ref{Sec:2}
we briefly discuss the background under consideration and describe
all the geometrical characteristics we need. In Sec. \ref{Sec:3}
we develop a general approach to calculation of the self-force in
this background for arbitrary profile of throat and apply it to
two specific kinds of profiles. We obtain simple formulae for the
self-force in these cases. However, this approach is valid only
for simple enough throat profiles, and therefore in Sec.
\ref{Sec:4} we develop an alternative method which is appropriate
for an arbitrary symmetric throat profile. We obtain general
formulae and find general expression for the self-force far from
the wormhole's throat. We discuss our results in Sec. \ref{Sec:5}.
Throughout this paper we use units $c=G=1$.

\section{The Background}\label{Sec:2}

Let us consider an asymptotically flat wormhole space-time. We
choose the line element of this space-time in the following form
\be
ds^2 = -dt^2 + d\rho^2 +r^2(\rho)(d\theta^2 + \sin^2 \theta
d\varphi^2),\label{ds^2}
\ee
where $t,\rho \in \mathds{R}$ and
$\theta \in [0,\pi], \varphi \in [0,2\pi]$. Profile of the
wormhole throat is described by the function $r(\rho)$. This
function must have a minimum at $\rho = 0$, and the minimal value
at $\rho = 0$ corresponds to the radius, $a$, of the wormhole
throat, \bd r(0) = a,\ \dot r(0) = 0, \ed where an over dot
denotes the derivative with respect to the radial coordinate
$\rho$. Space-time is naturally divided into two parts in
accordance with the sign of $\rho$. We shall label the part of the
space-time with positive (negative) $\rho$ and the functions on
this part with the sign "+"\ ("--").

The space-time possesses non-zero curvature. The scalar curvature
is given by
\bd
R = -\fr{2(2r\ddot r + \dot r^2 -1)}{r^2}.
\ed
Far from the wormhole throat the space-time becomes Minkowskian,
\be
r(\rho)|_{\rho \to \pm \infty} = \pm \rho.\label{condinfin}
\ee

Various kinds of throat profiles have been already considered in
another context \cite{KhuSus02}. The simplest model of a wormhole
is that with an infinitely short throat \cite{KhuSus02},
\be
r = a+|\rho|.
\ee
The space-time is flat everywhere except for the
throat, $\rho =0$, where the curvature has delta-like form,
\bd
R = -8\fr{\delta (\rho)}{a}.
\ed
Another wormhole space-time
that is characterized by the throat profile
\be
r = \sqrt{a^2+\rho^2}
\ee
is free of curvature singularities:
\bd
R = -\fr{2a^2}{(a^2+\rho^2)^2}.
\ed
The wormholes with the following profiles of throat
\bnn
r &=& \rho \coth \fr{\rho}b + a-b,\adb\\
r &=& \rho \tanh \fr{\rho}b + a,
\enn
have a throat whose length may be described using a parameter
$b$. The point is that for $\rho > b$ the space-time becomes
Minkowskian exponentially fast.

\section{Self-action in the Wormhole Space-time}\label{Sec:3}

Let us consider a charged particle at rest in the point
$\rho',\theta',\varphi'$ in the space-time with metric \Ref{ds^2}.
The Maxwell equation for zero component of the potential reads
\bd
\triangle A^0 = - \fr{4\pi e \delta (\rho - \rho')}{r^2(\rho)}
\fr{\delta (\theta - \theta') \delta (\varphi -
\varphi')}{\sin\theta}
\ed
where $\triangle = g^{kl} \nabla_k\nabla_l$. Due to static
character of the background we set other components of the vector
potential to be zero. It is obvious that $A^0 = 4\pi e
G(\mathbf{x};\mathbf{x'})$, where the three-dimensional Green's
function $G$ obeys the following equation
\bd
\triangle G(\mathbf{x};\mathbf{x'}) = - \fr{\delta (\rho -
\rho')}{r^2(\rho)} \fr{\delta (\theta - \theta') \delta (\varphi -
\varphi')}{\sin\theta}.
\ed
Due to spherical symmetry we may extract the angular dependence
(denoting succinctly $\Omega = (\theta,\varphi)$)
\bd
G(\mathbf{x};\mathbf{x'}) = \sum_{l=0}^\infty \sum_{m=-l}^l
Y_{lm}(\Omega) Y_{lm}^*(\Omega') g_l(\rho,\rho'),
\ed
and introduce the radial Green's function $g_l$ subject to the
equation
\be
\ddot g_l + \fr{2 r'}r \dot g_l - \fr{l(l+1)}{r^2}
g_l =- \fr{\delta (\rho - \rho')}{r^2(\rho)}.\label{maineq}
\ee
We represent the solution of this equation in the following form
\be
g_l = \theta(\rho-\rho') \Psi_1(\rho) \Psi_2 (\rho') +
\theta(\rho'-\rho) \Psi_1(\rho') \Psi_2 (\rho), \label{radialform}
\ee
where functions $\Psi$ are the solutions of the corresponding
homogeneous equation
\be
\ddot \Psi + \fr{2 r'}r \dot \Psi - \fr{l(l+1)}{r^2} \Psi =0,\label{radial}
\ee
satisfying the boundary conditions
\be
\lim_{\rho \to +\infty}\Psi_1 = 0,\ \lim_{\rho \to +\infty}\Psi_2 =
\infty.\label{condlimit}
\ee
If one substitutes \Ref{radialform} to \Ref{maineq} the condition
for the Wronskian emerges:
\be W(\Psi_1,\Psi_2) = \Psi_1 \dot \Psi_2 - \dot \Psi_1 \Psi_2 =
\fr 1{r^2(\rho)}.\label{wronskian1}
\ee

We consider the radial equation in domains $\rho > 0$ and $\rho <
0$ and obtain a pair of independent solutions $\phi^1,\phi^2$ for
each of the domains separately. We do not need to consider two
domains if it is possible to construct solutions that are
$C^1$-smooth over all space. However, this is not the case for
many situations. For the two kinds of the throat profile
considered below we may easily construct solutions for the two
domains separately (but not for all space). After that a procedure
developed here allows to construct $C^1$-smooth solution over all
space. Due to condition \Ref{condinfin} we obtain asymptotically
\bnn
\phi^1|_{\rho\to +\infty} &=& \rho^l,\adb\\
\phi^2|_{\rho\to +\infty} &=& \rho^{-l-1}.
\enn
The Wronskian of these solutions has the following form
\be
W(\phi^1_\pm,\phi^2_\pm) = \fr{A_\pm}{r^2(\rho)}\label{wronskian3}
\ee
with some constants $A_\pm$. Let us consider two different
solutions:
\bnn
\Psi_1 &=&\left\{ \begin{array}{lc}
                  \alpha^1_+ \phi^1_+ + \beta^1_+ \phi^2_+, & \rho >0 \\
                  \alpha^1_- \phi^1_- + \beta^1_- \phi^2_-, &
                  \rho <0
                \end{array} \right., \adb \\
\Psi_2 &=&\left\{ \begin{array}{lc}
                  \alpha^2_+ \phi^1_+ + \beta^2_+ \phi^2_+, & \rho >0 \\
                  \alpha^2_- \phi^1_- + \beta^2_- \phi^2_-, &
                  \rho <0
                \end{array} \right. .
\enn
The Wronskian condition \Ref{wronskian3} implies the
constraint on the coefficients: \bd \alpha^1_\pm \beta^2_\pm
-\beta^1_\pm \alpha^2_\pm = \fr 1{A_\pm}.
\ed

The general solution of the matching conditions, $\Psi_+(0) =
\Psi_-(0), \dot \Psi_+(0) = \dot \Psi_-(0)$, has the following
form
\bnn
\alpha_+ &=& \alpha_-
\left.\fr{W(\phi_-^1,\phi_+^2)}{W(\phi_+^1,\phi_+^2)}\right|_0 -
\beta_-
\left.\fr{W(\phi_+^2,\phi_-^2)}{W(\phi_+^1,\phi_+^2)}\right|_0, \adb \\
\beta_+ &=& \alpha_-
\left.\fr{W(\phi_+^1,\phi_-^1)}{W(\phi_+^1,\phi_+^2)}\right|_0 +
\beta_-
\left.\fr{W(\phi_+^1,\phi_-^2)}{W(\phi_+^1,\phi_+^2)}\right|_0,
\enn or vice versa \bnn \alpha_- &=& +\alpha_+
\left.\fr{W(\phi_+^1,\phi_-^2)}{W(\phi_-^1,\phi_-^2)}
\right|_0 + \beta_+ \left.\fr{W(\phi_+^2,\phi_-^2)}{W(\phi_-^1,\phi_-^2)}\right|_0,\adb \\
\beta_- &=& -\alpha_+
\left.\fr{W(\phi_+^1,\phi_-^1)}{W(\phi_-^1,\phi_-^2)} \right|_0 +
\beta_+ \left. \fr{W(\phi_-^1,\phi_+^2)}{W(\phi_-^1,\phi_-^2)}
\right|_0.
\enn
To satisfy the conditions \Ref{condlimit} we consider two specific
solutions that emerge when we set $\alpha^1_+=0$ and
$\alpha^2_-=0$. Corresponding solutions have the following form:
\bnn \Psi_1 &=&\left\{ \begin{array}{lc}
                  \beta^1_+ \phi^2_+, & \rho >0 \\
                  \alpha^1_- \phi^1_- + \beta^1_- \phi^2_-, &
                  \rho <0
                \end{array} \right., \adb \\
\Psi_2 &=&\left\{ \begin{array}{lc}
                  \alpha^2_+ \phi^1_+ + \beta^2_+ \phi^2_+, & \rho >0 \\
                  \beta^2_- \phi^2_-, &
                  \rho <0
                \end{array} \right. ,
\enn
where
\bd
\alpha^1_- = \beta^1_+
\left.\fr{W(\phi_+^2,\phi_-^2)}{W(\phi_-^1,\phi_-^2)}\right|_0,\
\beta^1_- = \beta^1_+ \left.
\fr{W(\phi_-^1,\phi_+^2)}{W(\phi_-^1,\phi_-^2)} \right|_0,
\ed
and
\bd
\alpha^2_+ = - \beta^2_-
\left.\fr{W(\phi_+^2,\phi_-^2)}{W(\phi_+^1,\phi_+^2)}\right|_0, \
\beta^2_+ = +\beta^2_-
\left.\fr{W(\phi_+^1,\phi_-^2)}{W(\phi_+^1,\phi_+^2)}\right|_0.
\ed
The Wronskian condition reads
\bd
-\beta^1_+ \alpha^2_+ = \fr 1{A_+},\ \alpha^1_- \beta^2_-  = \fr
1{A_-}.
\ed
Taking into account the above relations we obtain the radial
Green's function in the following form

\bs\label{g_lGen}
1. $\rho>\rho'>0$
\be\label{g^1}
g_l^{(1)}(\rho,\rho') =  -\fr 1{A_+}\phi^2_+(\rho')\phi^1_+(\rho) +
\left.\fr 1{A_+}\fr{W(\phi^1_+,\phi^2_-)}{W(\phi^2_+,\phi^2_-)}
\right|_0 \phi^2_+(\rho')\phi^2_+(\rho)
\ee

2. $0<\rho<\rho'$
\be
g_l^{(2)}(\rho,\rho') = g_l^{(1)}(\rho',\rho)
\ee

3. $\rho < \rho'$ and $\rho'>0,\ \rho<0$
\be
g_l^{(3)}(\rho,\rho') = \left.\fr
1{A_+}\fr{W(\phi^1_+,\phi^2_+)}{W(\phi^2_+,\phi^2_-)}
\right|_0\phi^2_+(\rho')\phi^2_-(\rho)
\ee

4. $\rho>\rho'$ and $\rho'<0,\ \rho>0$
\be
g_l^{(4)}(\rho,\rho')  = g_l^{(3)}(\rho',\rho')
\ee

5. $\rho' < \rho<0$
\be
g_l^{(5)}(\rho,\rho') = \fr 1{A_-}\phi^2_-(\rho')\phi^1_-(\rho) +
\left.\fr 1{A_-}\fr{W(\phi^1_-,\phi^2_+)}{W(\phi^2_+,\phi^2_-)}
\right|_0 \phi^2_-(\rho')\phi^2_-(\rho)
\ee

6. $\rho < \rho'<0$ \be g_l^{(6)}(\rho,\rho')  =
g_l^{(5)}(\rho',\rho) \ee \es where the constants $A_\pm $ may be
found from the relation \Ref{wronskian3}: \be A_\pm =
W(\phi^1_\pm,\phi^2_\pm)r^2(\rho)\label{wronskian6} \ee at
arbitrary point $\rho$.

Let us consider in detail the simple case of the symmetric throat
profile: $r(-\rho) = r(\rho)$. In this case we may choose
$\phi^{1,2}_-(\rho) = \phi^{1,2}_-(-\rho)$ and hence
$\phi^{1,2}_-(0) = \phi^{1,2}_-(0)$ and $\phi^{1,2}_-(0){}' =
-\phi^{1,2}_-(0){}'$ and $A_+ = -A_-$. Taking into account these
formulas we obtain

\bs\label{g_lGen3}
1. $\rho>\rho'>0$
\be
\label{g^1Gen3}
g_l^{(1)}(\rho,\rho') =  -\fr 1{A_+}\phi^2_+(\rho')\phi^1_+(\rho) +
\left.\fr 1{A_+}\fr{W_+(\phi^1_+,\phi^2_+)}{W_+(\phi^2_+,\phi^2_+)}
\right|_0 \phi^2_+(\rho')\phi^2_+(\rho)
\ee

2. $0<\rho<\rho'$
\be
g_l^{(2)}(\rho,\rho') = g_l^{(1)}(\rho',\rho)
\ee

3. $\rho < \rho'$ and $\rho'>0,\ \rho<0$
\be
g_l^{(3)}(\rho,\rho') = -\left.\fr
1{A_+}\fr{W(\phi^1_+,\phi^2_+)}{W_+(\phi^2_+,\phi^2_+)}
\right|_0\phi^2_+(\rho')\phi^2_+(-\rho)
\ee

4. $\rho>\rho'$ and $\rho'<0,\ \rho>0$
\be
g_l^{(4)}(\rho,\rho')  = g_l^{(3)}(\rho',\rho')
\ee

5. $\rho' < \rho<0$
\be
g_l^{(5)}(\rho,\rho') = g_l^{(1)}(-\rho,-\rho')
\ee

6. $\rho < \rho'<0$ \be g_l^{(6)}(\rho,\rho')  =
g_l^{(5)}(\rho',\rho) \ee
\es
Here $W_+(y_1,y_2)=y_1\dot y_2 + \dot y_1 y_2$. Thus we have to
write out in manifest form only $g_l^{(1)}$ and $g_l^{(3)}$.

The Green's function will not give a finite expression for the
self-force. The origin of this divergence is the electromagnetic
self-energy of point particle. There exist some approaches to
obtain finite result. First simple way is to consider total mass
as observed finite mass plus infinite electromagnetic
contribution. Usually this subtraction is called classical
"renormalization"\ because there is no Planck constant in the
divergent term. Dirac \cite{Dir38} suggested to consider radiative
Green's function to calculate the self-force. Since the radiative
Green's function is the difference between retarded and advanced
Green's functions, singular contribution cancels and we obtain a
finite result. There is also axiomatic approach suggested by Quinn
and Wald \cite{QuiWal97}. In the framework of this approach we
obtain finite expression by using a "comparison"\ axiom. This
approach was used in Refs. \cite{BurLiuSoe00,BurLiu01} for a
specific space-time.

We will use general approach to renormalization in curved space-time
\cite{BirDev} which means subtraction of the first terms from
DeWitt-Schwinger asymptotic expansion of a Green's function. In
general there are two kinds of divergences in this expansion,
namely, pole and logarithmic ones \cite{Chr78}. In three-dimensional
case which we are interested in there is only pole divergence, while
the logarithmic term is absent. The singular part of the Green's
function, which must be subtracted, has the following form (in $3D$
case)
\begin{displaymath}
 G^{sing} = \frac{1}{4\pi} \frac{\triangle^{1/2}}{\sqrt{2\sigma}},
\end{displaymath}
where $\sigma$ is half of the square of geodesic distance and
$\triangle$ is DeWitt-Morrett determinant. If we take coincidence
limit for angular variables then these quantities are easily
calculated using the metric (\ref{ds^2}): $\sigma = (\rho -
\rho')^2/2$ and $\triangle = 1$. Therefore to carry out
renormalization we have to subtract from the Green's function its
singular part, which has the following form:
\begin{displaymath}
 G^{sing} = \frac{1}{4\pi} \frac{1}{|\rho - \rho'|}.
\end{displaymath}
This is, in fact, the Green's function in Minkowski space-time.
This approach was used many times in different curved backgrounds
(e.g. see \cite{Khu05}). Now we are in a position to consider
first some specific cases in detail and after that to proceed to
the general profile of the throat.

\subsection{Profile $r = a + |\rho|$}

Two linearly independent solutions, $\phi^1,\phi^2$, are given by
\bs\label{phi}
\bn
\phi_\pm^{1} &=& r(\pm\rho)^{l} = (a\pm\rho)^{l},\adb \\
\phi_\pm^{2} &=& a^{2l+1}r(\pm\rho)^{-l-1} =
a^{2l+1}(a\pm\rho)^{-l-1},
\en
\es
with the Wronskian
\be
W(\phi_\pm^{1},\phi_\pm^{2}) = \mp(2l+1)
\fr{a^{2l+1}}{r^2(\rho)}.\label{wronskian2}
\ee

Therefore the solutions we need have the following form
\bnn
\Psi_1 &=&\beta^1_+ \left\{ \begin{array}{lc}
                  \phi^2_+, & \rho >0 \\
                  \fr{2(l+1)}{2l+1} \phi^1_- - \fr{1}{2l+1}\phi^2_-, &
                  \rho <0
                \end{array} \right., \adb \\
\Psi_2 &=&\beta^2_- \left\{ \begin{array}{lc}
                  \fr{2(l+1)}{2l+1} \phi^1_+ - \fr{1}{2l+1}\phi^2_+, & \rho >0 \\
                  \phi^2_-, & \rho <0
                \end{array} \right. .
\enn

From Eqs. \Ref{wronskian1} and \Ref{wronskian2} we obtain an
additional constraint
\bd
\beta^1_+ \beta^2_- = \fr 1{a^{2l+1}2(l+1)}.
\ed
Thus we have found the radial part of the Green's function in
manifest form
\bs
\bn
(2l+1)g_l^{(1)}(\rho,\rho') &=& \fr{r^l(\rho')}{r^{l+1}(\rho)} -
\fr{1}{2(l+1)}\fr{a^{2l+1}}{r^{l+1}(\rho)r^{l+1}(\rho')},\adb \\
(2l+1)g_l^{(3)}(\rho,\rho')  &=&
\fr{2l+1}{2(l+1)}\fr{a^{2l+1}}{r^{l+1}(\rho)r^{l+1}(\rho')}.
\en
\es

To calculate the full Green's function and the potential we turn
to the relation
\bd
\sum_{m=-l}^l Y_{lm}(\Omega)Y_{lm}^*(\Omega')
= \fr{2l+1}{4\pi} P_l(\cos\gamma),
\ed
by using which we obtain
\bd
G(x;x') = \fr 1{4\pi}\sum_{l=0}^\infty (2l+1) P_l(\cos\gamma)
g_l(\rho,\rho').
\ed
Here $\cos\gamma = \cos\theta\cos\theta' + \sin\theta\sin\theta'
\cos(\varphi-\varphi')$ is a cosine of an angle between two points
on a sphere. Now we use two series expansions:
\bnn
\sum_{k=0}^\infty t^k P_k(x) &=& \fr 1{\sqrt{1-2tx+t^2}},\adb\\
\sum_{k=0}^\infty \fr{t^{k+1}}{k+1} P_k(x) &=&
\ln\left|1+\fr{2t}{1-t+\sqrt{t^2-2tx+1}}\right|,
\enn
which allow us to obtain the Green's function in manifest form
\bnn
4\pi G^{(1)}(x;x') &=& \fr 1{\sqrt{r(\rho)^2
-2r(\rho)r(\rho')\cos\gamma + r(\rho')^2}} -
  \fr{1}{2a}\ln\left|1+\fr{2t}{1-t+\sqrt{t^2 -
  2t\cos\gamma+1}}\right|,\adb\\
4\pi G^{(3)}(x;x') &=& \fr{t}{a\sqrt{t^2 - 2t\cos\gamma+1}} -
  \fr{1}{2a}\ln\left|1+\fr{2t}{1-t+\sqrt{t^2 -
  2t\cos\gamma+1}}\right|,
\enn
where $t = \fr{a^2}{r(\rho) r(\rho')}$.

\begin{figure}[ht]
\begin{center}
\epsfxsize=9truecm\epsfbox{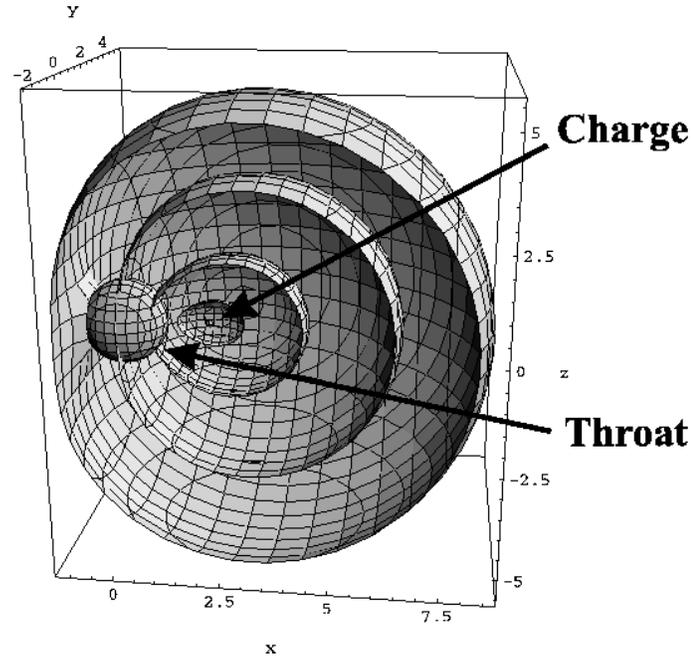}
\end{center}
\caption{Several surfaces of constant potential are shown. The
charge $e=1$ is at the point $x=2,y=z=0$. The small sphere is the
throat of the wormhole, $\rho=0$. We observe that some surfaces go
under the throat to another universe.} \label{Fig:p}
\end{figure}

Let us consider the case when the observation point, $\rho$, is
far from both the wormhole throat and from the position of a
particle, $\rho'$, that is we set $\rho \to \infty$. We consider
$\rho'>0$. In this limit we have
\bs\label{Afar}
\bn
\rho>0 &:& 4\pi G(x;x') = \fr 1{\rho} - \fr a{2\rho(a+\rho')}
+ O(\rho^{-2}),\\
\rho<0 &:& 4\pi G(x;x') = -\fr a{2\rho(a+\rho')} + O(\rho^{-2}).
\en
\es
From the first of these expressions we observe that the potential
of the particle contains an additional term with the same $\rho$
dependence alongside with the standard Coulomb part. The same
problem appears for a particle in the Schwarzschild space-time
\cite{Vil79}. To solve this problem it was suggested to add a
solution of the corresponding homogeneous equation to the solution
with a "bad"\ asymptotical behaviour \cite{Lin76}. However, in our
case there is no need for such modification: the potential
\Ref{Afar} correctly describes real potential of the particle at
rest. Indeed, let us consider the Gauss theorem in the space-time
of the wormhole. We place the particle at the point $\rho'$ and
enclose this charge by two spheres, namely, $S_1:\rho=R$ and
$S_2:\rho=-R$. The Gauss theorem reads
\bd
4\pi e = \iint_{S_1} \mathbf{En}dS + \iint_{S_2}
\mathbf{En}dS,
\ed
where $\mathbf{E}=-\nabla A_0$ and $n=(\pm 1,0,0)$ for $S_1$ and
$S_2$ correspondingly. Therefore we have
\be\label{Gauss}
4\pi e = -\iint_{S_1} \partial_\rho A_0|_{\rho=R}
dS + \iint_{S_2}
\partial_\rho A_0|_{\rho=-R} dS.
\ee
Taking into account formulae \Ref{Afar} it is easy to show
that the Gauss theorem is satisfied and we do not need to add a
solution of the homogeneous equation to correct the potential.
Thus the potential has "bad"\ behaviour at wide separations of the
observation point and the particle \bd A_0 = 4\pi e G(\rho;\rho')
\approx \fr e{\rho} \left[1- \fr a{2(\rho'+a)}\right] \ed instead
of the expected expression $e/\rho$. In fact, we observe the
particle with charge $e(1- \fr a{2(\rho'+a)})$ far from our
genuine charge $e$. The explanation is very easy -- some electric
field lines have gone through the throat to another, invisible
domain of space-time and for correct formulation of the Gauss
theorem we must take these field lines into account, too. Hence
the second sphere $S_2$ emerges. The Gauss theorem relates the
charge of the particle to the electric field flux density. In the
Fig. \ref{Fig:p} the surfaces of the constant potential $A_0$ of
the particle are plotted.

Now let us restrict ourselves to the $+$ part of space-time and
consider the situation when $\Omega=\Omega'$. We have the
following expression for the Green's function
\bd
G(\rho;\rho')=\fr{1}{4\pi} \fr{1}{|\rho-\rho'|} + \fr 1{8\pi a}
\ln [1-\fr{a^2}{r(\rho)r(\rho')}].
\ed
Therefore, after renormalization we obtain
\bd
G^{ren}(\rho;\rho')= \fr 1{8\pi a} \ln
[1-\fr{a^2}{r(\rho)r(\rho')}].
\ed
The potential at the position of the particle reads
\bd
\Phi=\fr e{2 a} \ln [1-\fr{a^2}{(a+\rho)^2}],
\ed
and the self-potential is
\be\label{U}
U=\fr{e^2}{4 a} \ln [1-\fr{a^2}{(a+\rho)^2}].
\ee
Some limiting cases are:
\bn
U|_{\rho\to 0} &=& \fr{e^2}{4a} \ln \fr{2\rho} a,\adb\nonumber \\
U|_{\rho\to\infty} &=& -
\fr{e^2a}{4\rho^2}+\fr{a^2}{2\rho^3},\adb\label{limit}
\\
U|_{a\to 0} &=& -\fr{a e^2}{4(a+\rho)^2}.\nonumber
\en
The self-force
\bd
\mathbf{F} = -\nabla U
\ed
has the radial component only
\bd
F^\rho = -\partial_\rho U = -\fr{ae^2}{2r(\rho)^3}\fr
1{1-\fr{a^2}{r(\rho)^2}}.
\ed
The self-force is always attractive, it turns into infinity at the
throat and goes down monotonically to zero as $\rho \to \infty$.
We may compare this expression with its analog for Schwarzschild
space-time with Schwarzschild radius $r_s = a$:
\bd
F^r = +\fr{ae^2}{2r^3}\sqrt{1-\fr{a^2}{r^2}}.
\ed
Important observations are:

1) The self-force in the wormhole space-time has an opposite sign
-- it is attractive.

2) Far from the wormhole throat and from the black hole we have
the same results but with opposite signs
\bnn
F^\rho_{wh} &=& -\fr{ae^2}{2\rho^3},\\
F^\rho_{bh} &=& +\fr{ae^2}{2\rho^3}.
\enn

3) At the Schwarzschild radius $r_s = a$ the self-force equals
zero, whereas at the wormhole throat it tends to infinity. The
latter discrepancy originates in the selected throat profile
function that leads to the curvature singularity at the throat.

\subsection{Profile $r = \sqrt{a^2+\rho^2}$}

The radial equation \Ref{radial} has the following two linearly
independent solutions
\bs\label{phi1}
\bn
\phi^{1}_+ &=& c_1^+P_l (z),\adb\label{phi1_1} \\
\phi^{2}_+ &=& c_2^+Q_l (z),\label{phi1_2}
\en
\es
with Wronskian
\be
W(\phi_+^{1},\phi_+^{2}) = ic_1^+c_2^+
\fr{a}{r^2}.\label{wronskian4}
\ee
Here $P_l$ and $Q_l$ are the Legendre polynomials of the first and
second kind, and $z=i\rho/a$. In the particular case when $c_1^+ =
\fr{(-i a)^l l!}{(2l-1)!!}, c_2^+ = \fr{(2l+1)!!
a^{l}}{(-i)^{l+1}l!}$ the solutions are real functions and far
enough from the wormhole we have
\bnn
\phi^1_+|_{\rho\to +\infty} &=& \rho^l,\adb\\
\phi^2_+|_{\rho\to +\infty} &=& a^{2l+1}\rho^{-l-1},
\enn
as in the above example. We should note that solutions given by
Eq. \Ref{phi1} are not $C^1$-smooth over all space. The point is
that the above functions have the following symmetry properties
\cite{BatErdV1}
\bd
P_l(-z) = (-1)^l P_l(z),\ Q_l(-z) = (-1)^{l+1} Q_l(z).
\ed
These conditions imply that the functions are not $C^1$-smooth at
the throat $\rho =0$. However, the procedure developed above
yields $C^1$-smooth solution provided that we have solutions in domains
$\rho>0$ and $\rho<0$.
For $\rho<0$ we choose
\bs\label{phi-}
\bn
\phi^{1}_- &=& c_1^-P_l (-z),\adb \\
\phi^{2}_- &=& c_2^-Q_l (-z),
\en
\es
with Wronskian
\be
W(\phi_-^{1},\phi_-^{2}) =
-ic_1^-c_2^-\fr{a}{r^2}.\label{wronskian5}
\ee

To simplify the expressions we take $c_1^- =c_1^+,\ c_2^- =
c_2^+$. Final result does not depend on the specific choice
because the formulae involve only the ratio of function values at
points $\rho$ and $\rho = 0$.

To proceed further we use the formulae (we take into account the
limit as $\rho\to 0$ along the imaginary axis from above)
\bnn
P_l(0) &=& \fr{\sqrt{\pi}}{\Gamma (\fr 12 - \fr l2)\Gamma (1+\fr
l2)},\ P_l'(0) = -\fr{2\sqrt{\pi}}{\Gamma (\fr 12 + \fr l2)\Gamma
(-\fr l2)},\\
Q_l(0) &=& \fr{\sqrt{\pi}}2 e^{-i\fr\pi 2 (l+1)} \fr{\Gamma (\fr 12
+ \fr l2)}{\Gamma (1+\fr l2)},\ Q_l'(0) = \sqrt{\pi} e^{-i\fr\pi 2
l} \fr{\Gamma (1 + \fr l2)}{\Gamma (\fr 12+\fr l2)}.
\enn
With the help of these relations it is easy to show that
\bn
\left.\fr{W_+(\phi^1_+,\phi^2_+)}{W_+(\phi^2_+,\phi^2_+)} \right|_0
&=& \fr{i}\pi \fr{c_1^+}{c_2^+},\label{W}\\
\left.\fr{W(\phi^1_+,\phi^2_+)}{W_+(\phi^2_+,\phi^2_+)} \right|_0
&=& (-1)^{l+1} \fr{i}\pi \fr{c_1^+}{c_2^-},\\
A_\pm &=& \pm ic_1^+ c_2^+ a.
\en

Now we use the formulae \Ref{g_lGen3} and \Ref{wronskian5} to
obtain the radial Green's function
\bn
g_l^{(1)}(\rho,\rho') &=& \fr ia P_l(z') Q_l(z) + \fr 1{\pi a} Q_l(z)Q_l(z'),\label{gPQ}\\
g_l^{(3)}(\rho,\rho') &=& \fr 1{\pi a}
(-1)^{l+1}Q_l(-z)Q_l(z').\nonumber
\en
Therefore
\bd
G^{(1)}(\rho,\rho') = \fr{1}{4\pi a}\sum_{l=0}^\infty (2l+1) \left\{
iP_l(z')Q_l(z) + \fr 1\pi Q_l(z')Q_l(z)\right\},
\ed
and
\bd
G^{(3)}(\rho,\rho') = \fr{1}{4\pi^2 a}\sum_{l=0}^\infty
(2l+1)(-1)^{l+1} Q_l(z')Q_l(-z),
\ed
where $z= i\rho/a$ and $z'=i\rho'/a$. Now we use the Heine formula
\cite{BatErdV1}
\bd
\sum_{l=0}^\infty (2l+1)P_l(z')Q_l(z) = \fr 1{z-z'}
\ed
and obtain
\bnn
4\pi G^{(1,2)}(\rho,\rho') &=& \fr 1{|\rho-\rho'|} + \fr{1}{\pi
a}\sum_{l=0}^\infty (2l+1)Q_l(z')Q_l(z),\\
4\pi G^{(3)}(\rho,\rho') &=& \fr{1}{\pi a}\sum_{l=0}^\infty
(2l+1)(-1)^{l+1}Q_l(z')Q_l(-z).
\enn

\begin{figure}[ht]
\begin{center}
\epsfxsize=9truecm\epsfbox{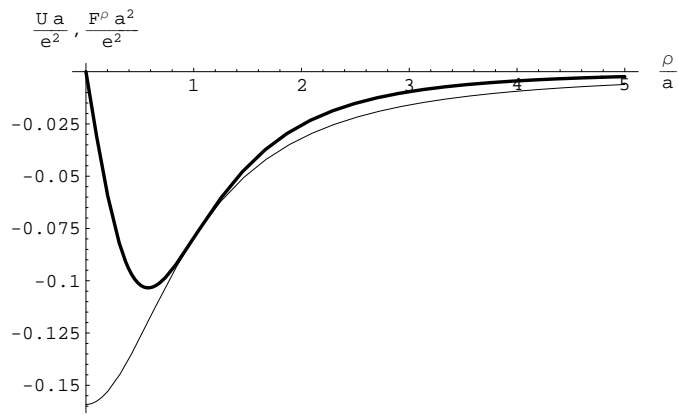}
\end{center} \caption{Thin line is plot of potential \Ref{U2} and
thick line is a
plot of the self-force \Ref{force}. The force has an extremum at
point $\rho^* = a/\sqrt{3}$.} \label{Fig:pf}
\end{figure}

To find these series in the closed form we use the integral
representation for the Legendre function of the second kind
\cite{BatErdV1}
\bd
Q_l (z) = \fr 12 \int_{-1}^1
\fr{P_l(t)}{z-t}dt, \ed which yields \bnn \sum_{l=0}^\infty
(2l+1)Q_l(ix')Q_l(ix) &=& -\fr{\arctan x - \arctan
x'}{x - x'},\adb \\
\sum_{l=0}^\infty (2l+1)(-1)^{l+1}Q_l(ix')Q_l(-ix) &=& -\fr{-\arctan
x + \arctan x' - \pi }{-x + x'}.
\enn
Therefore
\bnn
4\pi G^{(1,2)}(\rho,\rho') &=& \fr 1{|\rho-\rho'|} -\fr 1{\pi}
\fr{\arctan \fr{\rho}a - \arctan \fr{\rho'}a}{\rho - \rho'},\\
4\pi G^{(3)}(\rho,\rho') &=& -\fr 1{\pi}\fr{-\arctan \fr{\rho}a +
\arctan \fr{\rho'}a - \pi}{-\rho + \rho'}.
\enn
Far from the wormhole throat we have
\bnn
\lim_{\rho\to +\infty}4\pi G^{(1,2)}(\rho,\rho') &=& \fr 1{\rho} +
\fr 1{\rho}\left[-\fr 12 + \fr 1{\pi}
\arctan \fr{\rho'}a \right],\\
\lim_{\rho\to -\infty}4\pi G^{(3)}(\rho,\rho') &=& \fr
1{\rho}\left[-\fr {1}2 + \fr 1\pi \arctan \fr{\rho'}a\right],
\enn
hence the potential far from the charge kept at the point $\rho'$
has the following form
\bnn
A^0_{(1)} &=& \fr e{\rho} + \fr e{\rho}\left[-\fr 12 + \fr 1{\pi}
\arctan \fr{\rho'}a \right],\\
A^0_{(3)} &=& \fr e{\rho}\left[-\fr {1}2 + \fr 1\pi \arctan
\fr{\rho'}a\right],
\enn
and again it obeys the Gauss theorem \Ref{Gauss}. The renormalized
Green's function is given by
\bnn
4\pi G(\rho,\rho)^{ren} &=& \fr{1}{\pi a}\sum_{l=0}^\infty
(2l+1)Q_l(z)^2\adb \\
&=& -\fr{1}{\pi} \fr a{\rho^2 +a^2}.
\enn
Therefore we obtain the self-potential
\be\label{U2}
U =
-\fr{e^2}{2\pi} \fr a{\rho^2 +a^2}
\ee
and the self-force
\be
F^\rho = \partial_\rho U = -\fr{e^2}{\pi}
\fr{a\rho}{(\rho^2 +a^2)^2}.\label{force}
\ee
As expected the self-force is everywhere finite and equals zero at
the throat. Far from the wormhole we have
\bd
F^\rho \approx
-\fr{e^2}{\pi} \fr{a}{\rho^3}.
\ed
Thus the self-force is always attractive. It has maximum value at
distance $\rho^*=a/\sqrt{3}$ with magnitude $F^\rho_{max} =
3\sqrt{3}e^2/16\pi a^2$. The plots of the potential and the
self-force are shown in the Fig. \ref{Fig:pf}.

\section{General case}\label{Sec:4}
In this section we outline our approach to generalization of the
results of the previous sections. From now on the profile function
$r(\rho)$ will be considered to be arbitrary symmetric function
(the condition of asymptotic flatness given by Eq.
(\ref{condinfin}) is understood).

Let us perform the WKB analysis of the defining equation for
radial Green's functions
\be
\label{radial_}
\ddot\phi + \fr{2 \dot r}r \dot\phi - \fr{l(l+1)}{r^2} \phi = 0.
\ee
To accomplish this we represent the solution of this equation
in the form
\be
\phi = e^{S},\label{e^S}
\ee
and arrive at the following equation for $S$:
\be
\ddot S+\dot S^2 + \fr{2\dot r}r \dot S - \fr{\nu^2-1/4}{r^2}
=0,\label{S}
\ee
where $\nu = l+1/2$. Next step is to expand $S$ in the following
power series in $\nu$ \cite{Khu03}:
\be
S = \sum_{n=-1}^\infty \nu^{-n} S_n.\label{Sexp}
\ee

One comment is in order. Equation \Ref{S} is valid for arbitrary
$l \geq 0$, whereas the expansion over $1/\nu$ \Ref{Sexp} is
fulfilled for $l>0$ only. For $l=0$ we have $1/\nu = 2$ and the
expansion is divergent. Nevertheless, we may easily find $g_l$ for
$l=0$ without turning to the series expansion. Indeed, for $l=0$
the equation \Ref{radial_} simplifies considerably ($\varphi$
stands here for the zero mode only):
\bd
\ddot\varphi + \fr{2 \dot r}r \dot\varphi = 0.
\ed
This equation has two independent solutions:
\bd
\varphi_+^1 = C_1,\ \varphi_+^2 = C_2\int_{\rho_0}^\rho
\fr{1}{r^2}d\rho + C_3.
\ed
We take $C_1 = 1$ in order to be in agreement with \Ref{phi} but this is
insufficient. This solution corresponds to the solution $\sim\rho^l$ for $l=0$.
The second solution must tend to zero far from the throat and it
corresponds to the solution $\sim\rho^{-l-1}$ for $l=0$. Therefore
we may write
\bd
\lim_{\rho\to\infty}\varphi_+^2(\rho) = C_2\int_{\rho_0}^\infty
\fr{1}{r^2}d\rho + C_3 = 0,
\ed
hence
\bd
C_3 = -C_2\int_{\rho_0}^\infty \fr{1}{r^2}d\rho.
\ed
Thus we obtain the second solution in the form
\bd
\varphi_+^2(\rho) = -C_2\int_{\rho}^\infty \fr{1}{r^2}d\rho.
\ed
We set $C_2 = -a$ to be in agreement with \Ref{phi}. To summarize,
we have
\be
\varphi_+^1 = 1,\ \varphi_+^2 = \int_\rho^\infty \fr{a}{r^2}d\rho,
\label{phizero}
\ee
with Wronskian $W(\varphi_+^1,\varphi_+^2) = -a/r^2$. These
expressions allow us to obtain zero component of the Green's
function
\bd
g_0^{(1)}(\rho,\rho') = \fr 1a \varphi^2_+(\rho) - \fr 1{2a}
\varphi^2_+(\rho)\fr{\varphi^2_+(\rho')}{\varphi^2_+(0)}.
\ed

Now that we have separated the $l=0$ case, it is possible to
proceed with $l>0$ contributions, using the series expansion.
Substituting \Ref{Sexp} to \Ref{S} yields the following set of
recurring expressions for the functions $S_n$:

\bnn
\dot S_{-1} &=& \pm \fr 1 r,\adb\\
\dot S_0    &=& - \fr{\ddot S_{-1}}{2\dot S_{-1}} -
\fr{\dot r}{r},\adb \\
\dot S_1 &=& - \fr 1{2\dot S_{-1}} \left[ \ddot S_0 + \dot S_0^2 +
\fr{2\dot r}{r} \dot S_0 + \fr 1{4r^2}\right],\adb \\
\dot S_{m+1} &=&- \fr 1{2\dot S_{-1}} \left[ \ddot S_m +
\sum_{n=0}^m \dot S_n \dot S_{m-n} + \fr{2\dot r}{r} \dot
S_m\right],\ m=1,2,\ldots .
\enn
The sign ``$+$'' in the first equation corresponds to the solution
$\phi^1$ and sign ``$-$'' to the solution $\phi^2$ which tends to zero
as $\rho \rightarrow \infty$. Following are the manifest
expressions for derivatives of the first several functions $S_n$:
\bnn
\dot S_{-1} &=& \pm \fr 1 r,\adb\\
\dot S_0    &=& - \fr{\dot r}{2r} = -\fr 12 (\ln r)^\cdot,\adb \\
\dot S_1 &=& \pm\fr{ 1}{4} \left[ \ddot r + \fr{\dot r^2}{2r} -
\fr{1}{2r}\right],\adb \\
\dot S_2 &=& -\fr 18 \left[r r^{(3)} + 2 \dot r \ddot r\right]= - \fr 18 (r \ddot r + \fr 1{2} \dot r^2)^\cdot, \adb\\
\dot S_3 &=& \pm \fr 1{16}\left[ r^2 r^{(4)} + 4 r\dot r r^{(3)} +
\fr 32 r \ddot r^2 + \fr 32 \dot r^2 \ddot r + \fr 12 \ddot r
-\fr{\dot r^4}{8 r} + \fr{\dot r^2}{4r} - \fr 1{8r}\right],\adb \\
\dot S_4 &=& -\fr 1{32} \left[r^3 r^{(5)} + 7 r^2 \dot r r^{(4)} + 6
r^2 \ddot r r^{(3)} + 9 r \dot r^2 r^{(3)} + r r^{(3)} + 4 r \dot r
\ddot r^2 + 2 \dot r \ddot r \right]\adb\\
&=& -\fr 1{128} \left( 4r^3 r^{(4)} + 16 r^2 \dot r r^{(3)} +4 r^2
\ddot r^2 + 4 r \dot r^2 \ddot r + 4 r \ddot r -\dot r^4 + 2 \dot
r^2\right)^\cdot
\enn
Note that all $\dot S_{2n}$ come with only one sign, whereas the
signs of $\dot S_{2n+1}$ vary according to the choice between
$\phi^1$ and $\phi^2$. Another important point is that all $\dot
S_{2n}$ appear to be full derivatives, in contrast to all $\dot
S_{2n+1}$. These two points will be essential in the subsequent
analysis.

A simple verification of the approach can be done at this point.
Taking $r = \sqrt{\rho^2+a^2}$ we can evaluate the expansion
\Ref{Sexp} and then compare \Ref{e^S} with the exact solutions
\Ref{phi1}. Numerical calculation reveals fast convergence of the
series to the exact solutions. The series that results when all
$\dot S_{2n+1}$ are taken with the ``$-$'' sign coincides with the
solution $i^{n+1}Q_l(z)$ \Ref{phi1_2}, while the series with the
``$+$'' sign corresponds to the following linear combination of
\Ref{phi1_1} and \Ref{phi1_2}: $(-i)^n (P_l(z) - \fr i\pi
Q_l(z))$. We would like to note that the result of the previous
section would be the same if we had used this linear combination
instead of $(-i)^n P_l(z)$.

Let us introduce a pair of indices to help us distinguish between
different solutions. From now on $S^1$ will denote the whole
series \Ref{Sexp}, in which all $\dot S_{2n+1}$ are taken with the
``$+$'' sign, while $S^2$ will correspond to the ``$-$'' sign. These are
essentially the two different solutions that we previously called
$\phi^1$ and $\phi^2$. We will additionally mark the solutions
with ``$\pm$'' according to the sign of $\rho$. The Wronskian
condition \Ref{wronskian3} is therefore cast into the form

\bd
e^{S^{\pm 1}(\rho) + S^{\pm 2}(\rho)} = \fr{A_\pm}{r^2(\rho)} \fr
1{\dot S^{\pm 2}(\rho) - \dot S^{\pm 1}(\rho)},
\ed
and evaluating it at $\rho=0$ we get
\bd
e^{S^{\pm 1}(0) + S^{\pm 2}(0)} = \fr{A_\pm}{a^2} \fr 1{\dot S^{\pm
2}(0) - \dot S^{\pm 1}(0)}.
\ed
Functions $\dot S_{2n}$ with even indices have the following
structure
\bd
\dot S_{2n}(\rho) = \sum_{k=0}^n s_{2k+1}r^{(2 k+1)};
\ed
each term necessarily contains odd order derivatives of $r$ with
respect to $\rho$. Hence if the profile function $r(\rho)$ is
symmetric, $r(-\rho) = r(\rho)$, then $\dot S_{2n}(0) = 0$.

By using the above formulae we may represent the Green's function
\Ref{g^1Gen3} in the following form
\bn
\label{g_l}
g_l^{(1)}(\rho,\rho') &=& -\fr 1{a^2} \fr{e^{S^{+2}(\rho) -S^{+2}(0) +
S^{+1}(\rho') -S^{+1}(0)}}{\dot S^{+2}(0)-\dot S^{+1}(0)}\\
&+&\fr 1{a^2} \fr{e^{S^{+2}(\rho) -S^{+2}(0) + S^{+2}(\rho')
-S^{+2}(0)}}{\dot S^{+2}(0)-\dot S^{+1}(0)} \fr{\dot S^{-2}(0)-\dot
S^{+1}(0)}{\dot S^{-2}(0)-\dot S^{+2}(0)}.\nonumber
\en
Due to symmetry of the profile function, $r(-\rho) = r(\rho)$,
we may replace $\dot S^{-2}(0) = - \dot S^{+2}(0)$. From the above
chain we obtain in manifest form:
\bnn
S^{1,2}_{-1}(\rho)-S^{1,2}_{-1}(0) &=& \pm \int_0^\rho \fr{d\rho}{r(\rho)},\\
S^{1,2}_{0}(\rho)-S^{1,2}_{0}(0) &=& \fr 12 \ln \fr a{r(\rho)}.
\enn
Therefore the radial Green's function reads $(l>0)$
\bnn
g_l(\rho,\rho') &=& \fr 1{2\nu} \fr{e^{-\nu \int_{\rho'}^\rho
\fr{d\rho}{r}}}{\sqrt{r(\rho) r(\rho')}}
 \fr{e^{\sum_{n=1}^\infty \nu^{-n}(S^{+2}_n(\rho) +
S^{+1}_n(\rho'))}}{\sum_{n=0}^\infty
\nu^{-2n} (a \dot S^{+1}_{2n-1}(0))}\adb \\
&-& \fr 1{4\nu} \fr{e^{-\nu \left[\int_{0}^\rho \fr{d\rho}{r} +
\int_{0}^{\rho'} \fr{d\rho}{r}\right]}}{\sqrt{r(\rho) r(\rho')}}
 \fr{e^{\sum_{n=1}^\infty \nu^{-n}(S^{+2}_n(\rho) +
S^{+2}_n(\rho'))}}{\sum_{n=0}^\infty \nu^{-2n} (a \dot
S^{+1}_{2n-1}(0))} \fr{\sum_{n=0}^\infty \nu^{-2n} (-2 a \dot
S^{+1}_{2n}(0))}{\sum_{n=0}^\infty \nu^{-n+1} (-a \dot
S^{+2}_{n-1}(0))}.
\enn
The second part here gives zero contribution because $\dot
S_{2n}^{+1}$ contains only odd order derivatives of $r$, which are
zero at $\rho=0$. This fact may be verified straightforwardly by
performing calculations for the $\sqrt{\rho^2+a^2}$ profile
function. In this case
\bd
\dot S^{+2}(0) + \dot S^{+1}(0) = 0.
\ed
We must take into account that $S^{+2}$ corresponds to $P_l(z)$
and $S^{+1}$ corresponds to $P_l(z)-\fr i\pi Q_l(z)$. This issue
was already addressed above.

Let us now perform summation over $l$ in order to obtain full
Green's function. We have
\bnn
\sum_{l=0}^\infty \fr {2\nu}{4\pi}g_l(\rho,\rho') &=& \fr
1{4\pi}\sum_{l=1}^\infty \fr{e^{-\nu \int_{\rho'}^\rho
\fr{d\rho}{r}}}{\sqrt{r(\rho) r(\rho')}}
 \fr{e^{\sum_{n=1}^\infty \nu^{-n}(S^{+2}_n(\rho) +
S^{+1}_n(\rho'))}}{\sum_{n=0}^\infty \nu^{-2n} (a \dot
S^{+1}_{2n-1}(0))}+ \fr 1{4\pi} g_0(\rho,\rho')\adb \\
&=&\fr 1{4\pi} \fr{1}{\sqrt{r(\rho) r(\rho')}}\sum_{k=0}^\infty
f_k(b) j_k (\rho,\rho')+\fr 1{4\pi} g_0(\rho,\rho'),
\enn
where
\bnn f_k(b) &=& \sum_{l=1}^\infty \fr 1{\nu^k} e^{-\nu b}=
\sum_{l=0}^\infty \fr 1{\nu^k} e^{-\nu b}- 2^k e^{-\fr 12 b},\adb\\
b &=& \int_{\rho'}^\rho \fr{d\rho}{r(\rho)}.
\enn
To obtain the above formula we have permuted the summations over
$k$ and over $l$. The functions $f_k$ are expressed in terms of
the function $\Phi$ \cite{BatErdV1}
\bd
f_k(b) = e^{-\fr b2} \Phi (e^{-b},k,\fr 32).
\ed
First two functions in manifest form are
\bnn
f_0(b) &=& \fr 12 \fr 1{\sinh(\fr b2)}- e^{-\fr 12 b},\adb \\
f_1(b) &=& \ln \fr{1+e^{-\fr b2}}{1-e^{-\fr b2}}- 2 e^{-\fr 12 b}.
\enn

To calculate the self-force we need the coincidence limit
$\rho'\to\rho$. In this case only first two function $f_k$ are
divergent
\bnn
\fr{f_0(b)}{\sqrt{r(\rho)r(\rho')}} &=& \fr 1{\rho-\rho'} - \fr 1r + O(\rho-\rho'),\adb \\
\fr{f_1(b)}{\sqrt{r(\rho)r(\rho')}} &=& -\fr 1 r \ln \fr{\rho
-\rho'}{4r} - \fr 2r + O(\rho-\rho'),\adb \\
\fr{f_k(b)}{\sqrt{r(\rho)r(\rho')}} &=& \fr 1r \zeta_H(k,\fr 32) +
O(\rho-\rho'),
\enn
where $\zeta_H$ is the Hurwitz zeta-function \cite{BatErdV1}.

The functions $j_k$ can not be found in closed form for arbitrary
index $k$, however, each function may be found from the expansion
over $\nu$. In manifest form we obtain the following expressions
for these functions:
\bnn
j_0 (\rho,\rho') &=& 1,\adb \\
j_1 (\rho,\rho') &=& -\int_{\rho'}^\rho \fr{-1+\dot r^2 + 2r\ddot
r}{8r} d\rho = - \fr{-1+\dot r^2 + 2r\ddot r}{8r}(\rho-\rho') +
O((\rho-\rho')^2),\adb\\
j_2 (\rho,\rho) &=& - \fr{-1+\dot r^2 + 2r\ddot r}{8}= \fr{3}{8} a_1r^2,\adb\\
j_4 (\rho,\rho) &=& \fr 1{128} [3+3\dot r^4 - 12 r\ddot r - 4r^2
\ddot r^2 -2 \dot r^2 (3+2r\ddot r) - 32 r^2 \dot r r^{(3)} - 8 r^3
r^{(4)}],\adb\\
j_6 (\rho,\rho) &=& \fr{1}{1024}[5 - 5\dot r^6 - 30r\ddot r - 20r^2
\ddot r^2 + 3\dot r^4(5 + 6r\ddot r) - 192r^2\dot r^3
r^{(3)}\adb\\
&-& 8r^3(\ddot r^3 + 5r^{(4)}) - \dot r^2 (15 + 20r \ddot r + 12r^2
\ddot r^2 + 456r^3 r^{(4)})\adb \\
&-& 8r^4 (11r^{(3)}{}^2 + 20 \ddot rr^{(4)}) - 16r^2 \dot r ( 10 ( 1
+ 3 r \ddot r) r^{(3)} + 11 r^2 r^{(5)}) - 16 r^5 r^{(6)}].
\enn
In the above expressions $a_1$ stands for the first heat kernel
coefficient (see review \cite{Vass03}).

All terms of the form $f_{2k+1}j_{2k+1}$ vanish in the coincidence
limit. Therefore we obtain
\bnn
\sum_{l=0}^\infty \fr {2\nu}{4\pi}g_l(\rho,\rho') &=& \fr 1{4\pi}
\left[\fr 1{\rho-\rho'} - \fr 1r + \fr 1r \sum_{k=1}^\infty
\zeta_H(2k,\fr 32) j_{2k} (\rho,\rho)+ \fr 1a \varphi^2_+(\rho) -
\fr 1{2a}
\varphi^2_+(\rho)\fr{\varphi^2_+(\rho)}{\varphi^2_+(0)}\right].
\enn
After regularization we arrive at the formula
\be
\label{U_renorm'd}
U(\rho) = \fr{e^2}{2} \left[- \fr 1r + \fr 1r \sum_{k=1}^\infty
\zeta_H(2k,\fr 32) j_{2k} (\rho,\rho)+ \fr 1a \varphi^2_+(\rho) -
\fr 1{2a}
\varphi^2_+(\rho)\fr{\varphi^2_+(\rho)}{\varphi^2_+(0)}\right],
\ee
This expression is exact and we may use it for arbitrary throat
profile.
\begin{figure}[ht]
\begin{center}
\epsfxsize=9truecm\epsfbox{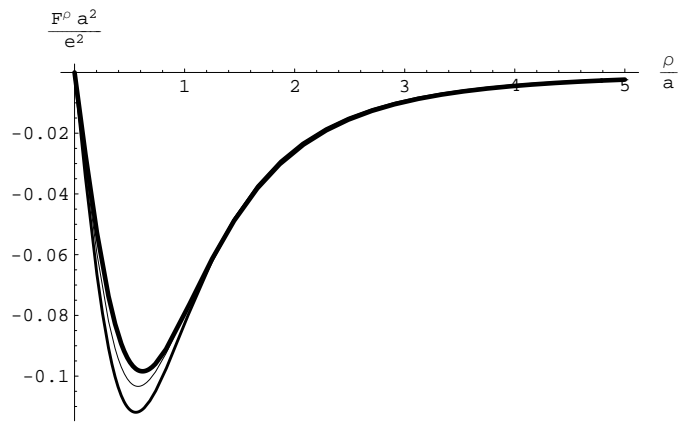}
\end{center}
\caption{Thin line is the plot of the exact formula for the
self-force given by Eq. \Ref{force}. Middle thickness line is the
plot of the self-force calculated by the above formula up to the
first term $k=1$. Thick line is the same formula up to $k=2$.}
\label{Fig:f}
\end{figure}

Fig. \ref{Fig:f} shows the exact form of the self-force for the
second kind of profile of the throat and the result obtained from
the above formula. We observe that even the first term of the
series gives good approximation. In fact both expressions are very
close for small and large distances from the wormhole throat,
there is small deviation close to the extremum only. For this
reason we may write out an approximate expression for
self-potential in which there is contribution from first heat
kernel coefficient
\be
U(\rho) \approx \fr{e^2}{2} \left[- \fr 1r + \zeta_H(2,\fr 32) \fr
38 a_1 r^2 + \fr 1a \varphi^2_+(\rho) - \fr 1{2a}
\varphi^2_+(\rho)\fr{\varphi^2_+(\rho)}{\varphi^2_+(0)}\right],
\ee
where $\zeta_H(2,\fr 32) = \fr{\pi^2}2 -4 \approx 0.9348$.

We may obtain general conclusion from \Ref{U_renorm'd}. Let us
consider large distances from the throat and suppose that the profile function
has the following asymptotic series expansion:
\bd
r = \rho + \sum_{n=0}^\infty v_n \rho^{-n}.
\ed
Each coefficient $j_{2k}(\rho,\rho)$ has the same asymptotic form
\bd
j_{2k}(\rho,\rho) = - \fr{v_1}{4\rho^2} + O(\rho^{-3}).
\ed
The relation
\bd
\sum_{k=1}^\infty \zeta_H(2k,\fr 32) = \fr 43
\ed
yields
\bd
\fr 1r \sum_{k=1}^\infty \zeta_H(2k,\fr 32) j_{2k} (\rho,\rho) = -
\fr{v_1}{3\rho^3} + O(\rho^{-4})
\ed
Putting all expressions together we obtain the following
asymptotic expansion for the renormalized Green's function
\bd
G^{ren}(\rho;\rho) = -\fr 1{8\pi\rho^2}\fr a{\varphi^2_+(0)} + \fr
1{4\pi\rho^3}\fr{a v_0}{\varphi^2_+(0)} + O(\rho^{-4}),
\ed
and for the self-potential
\bd
U = -\fr {e^2}{4\rho^2}\fr a{\varphi^2_+(0)} +
\fr{e^2}{2\rho^3}\fr{a v_0}{\varphi^2_+(0)} + O(\rho^{-4}).
\ed
Let us compare this result with the cases considered above. For
the profile $r = a + |\rho|$ we have $v^0 = a$ and
\bd
\varphi^2_+(0) = \int_0^\infty \fr{ad\rho}{(a+\rho)^2}  = 1.
\ed
Therefore
\bd
U = -\fr {ae^2}{4\rho^2} + \fr{a^2e^2}{2\rho^3} + O(\rho^{-4}).
\ed
This expression coincides with the result that follows
from the exact formula \Ref{limit}. For the smooth throat profile
$\sqrt{\rho^2+a^2}$ we have $v^0 =0$ and
\bd
\varphi^2_+(0) = \int_0^\infty \fr{ad\rho}{a^2+\rho^2}  = \fr \pi 2.
\ed
Therefore
\bd
U = -\fr {ae^2}{2\pi \rho^2} + O(\rho^{-4})
\ed
in agreement with Eq. \Ref{U2}.

Thus we may conclude that far from the wormhole throat we have the
following expression for the self-potential
\bd
U = -\fr {e^2}{4\rho^2}\fr a{\varphi^2_+(0)} = -\fr {e^2}{4\rho^2}
\left[\int_0^\infty \fr{d\rho}{r^2(\rho)}\right]^{-1}.
\ed
Note that it is always negative, hence the self-force is an
attractive force. All information about the specific throat
profile is encoded in the factor
\bd
\int_0^\infty \fr{d\rho}{r^2(\rho)}.
\ed

\section{Conclusion}\label{Sec:5}

The aim of the paper was to calculate the self-force acting on a
point charge at rest in the space-time of a wormhole. This
calculation has relevance to many problems. First of all the
space-time of a wormhole is an example of topologically
non-trivial space-time and it is interesting to clear up the role
of non-trivial topology of the wormhole space-time. As far as we
know this is the first calculation of this kind. The second point
is that the self-force may play a serious role in wormhole
physics.

We have developed standard approach to calculation of the
self-force. Following it we have obtained the self-force
explicitly for two profiles of the wormhole's throat. The first
shape of the throat is characterized by a curvature singularity --
the space-time is everywhere flat except for the throat. The
self-force is everywhere attractive and singular at the throat.
The singularity originates in the accepted model of the throat.
Far from the wormhole's throat the self-force has the same
magnitude as in the Schwarzschild space-time with the
Schwarzschild radius equal to the throat radius, but the sign of
the force is opposite. In the second example we have carried out
the calculations for a smooth throat profile. The self-force is
zero at the throat, as expected due to symmetry of the space-time
with respect to the throat. In both examples we observe that the
self-force is attractive, irrespective of the particle position.
We suppose that this is inherent to the wormhole with an arbitrary
throat profile and is the manifestation of non-trivial topological
structure of space-time. The manifest calculations of self-force
for arbitrary throat profile developed above show that far from
the throat the self-force is always attractive and falls down as
third power of distance. The magnitude of this force depends on
the throat profile.

Let us now speculate about the result obtained. Charged particle
in the wormhole space-time is always attracted towards the
wormhole's throat. Starting from rest at infinity it will reach
the throat with kinetic energy (for the second kind of throat
profile considered)
\bd
\fr{mv^2}{2}=\fr{e^2}{2\pi a} = -U_{\rho = 0}.
\ed

Then it goes through the throat to another universe. The
self-force will still attract the particle to the throat and the
velocity of the particle will decrease. The particle will finally
stop and the process will repeat. The process is similar to the
oscillations of a particle in the potential well. Due to the
acceleration the particle will lose its energy for radiation and
in the end it will be at rest at the throat. Therefore particles
will concentrate at the throat, at the minimum of the self-energy.
The intensity of this process is managed by the characteristic
value of the self-force, $e^2/a^2$, which is smaller for
macroscopic wormholes. Therefore the smaller wormhole's throat,
the faster it will gather neighbouring charged particles. At the
same time particles at the throat with the same sign of charge
will repel each other by Coulomb force. The magnitude of the force
is defined by the same parameter $e^2/a^2$. Thus an equilibrium
configuration of the particles near the throat is likely to exist.
We can not exploit this picture for uncharged particles. Usually,
the self-force for massive uncharged particle has opposite sign
comparing with electromagnetic charge. The magnitude of the force
is much smaller and defined by parameter $Gm^2$ instead of $e^2$
in the electromagnetic case. With reference to our case it means
that the uncharged massive particle might be repelled by the
wormhole.

\begin{acknowledgments} NK would like to thank B. Linet and S. Krasnikov for
helpful comments on the paper. This work was supported by the Russian
Foundation for Basic Research Grants No. 05-02-17344-a and 05-02-39023-a.
\end{acknowledgments}


\begin{thebibliography}{99}
\bibitem{EinRos35} A. Einstein and N. Rosen, Phys. Rev. \textbf{48},
73 (1935).
\bibitem{WheBook} J.A. Wheeler, Phys. Rev. \textbf{97}, 511 (1955).
\bibitem{MorTho88} M.S. Morris and K.S. Thorne, Am. J. Phys. \textbf{56}, 395
(1988).
\bibitem{MorThoYur88} M.S. Morris, K.S. Thorne and U. Yurtsever, Phys. Rev.
Lett. \textbf{61}, 1446 (1988).
\bibitem{VisBook} M. Visser, \textit{Lorentzian Wormholes: From Einstein to
Hawking}, American Institute of Physics, Woodbury, NY, 1995.
\bibitem{Kra00} S. Krasnikov, Phys. Rev. \textbf{D62}, 084028
(2000).
\bibitem{KhuSus02} N.R. Khusnutdinov and S.V. Sushkov, Phys. Rev. \textbf{D65},
084028 (2002); N.R. Khusnutdinov, Phys. Rev. \textbf{D67}, 124020 (2003).
\bibitem{Gar05} R. Garattini, Class. Quant. Grav. \textbf{22}, 1105 (2005).
\bibitem{Arm02} C. Armendariz-Picon, Phys. Rev. \textbf{D65} 104010 (2002).
\bibitem{Sus05} S. Sushkov, Phys. Rev. \textbf{D71}, 043520 (2005).
\bibitem{Lob05} F.S.N. Lobo, Phys. Rev. \textbf{D71}, 084011 (2005).
\bibitem{Lob05_2} F.S.N. Lobo, Phys. Rev. \textbf{D71}, 124022 (2005).
\bibitem{DamSol07} T. Damour, S.N. Solodukhin, Phys. Rev. \textbf{D76}, 024016
(2007).
\bibitem{Nar05} R. Narayan, New J. Phys. \textbf{7}, 199 (2005).
\bibitem{SchLeiRob06} R.E. Schild, D.J. Leiter, and S.L. Robertson,
\emph{Observations supporting the existence of an intrinsic magnetic
moment inside the central compact object within the quasar
Q0957+561}, astroph/0505518.
\bibitem{KarNovSha06} N.S. Kardashev, I.D. Novikov and A.A. Shatskiy,
Int. J. Mod. Phys. \textbf{D16}, 909 (2007).
\bibitem{DeWBre60} B.S. DeWitt, R.W. Brehme, Ann. Phys. \textbf{9}, 220 (1960).
\bibitem{Poi} E. Poisson,  Living Reviews in Relativity \textbf{7}, 6 (2004).
\bibitem{Khu05} N.R. Khusnutdinov, Physics - Uspekhi \textbf{48}, 577
(2005)[Uspekhi Fizicheskikh Nauk \textbf{175}, 603 (2005)].
\bibitem{Lin86} B. Linet, Phys. Rev. \textbf{D33}, 1833 (1986).
\bibitem{Gal90} D.V. Gal'tsov, Fortschr. Phys. \textbf{38}, 945 (1990).
\bibitem{KhuBez01} N.R. Khusnutdinov and V.B. Bezerra, Phys. Rev.
\textbf{D64}, 083506 (2001).
\bibitem{KhaKhuSus06} A.R. Khabibullin, N.R. Khusnutdinov and
S.V. Sushkov, Class. Quantum Grav. \textbf{23}, 627 (2006).
\bibitem{BirDev} N.D. Birrell and P.C.W. Davies, \textit{Quantum Fields
in Curved Space} Cambridge University Press, Cambridge, England,
1982.
\bibitem{Chr78} S.M. Christensen, Phys. Rev. \textbf{D17}, 946 (1978).
\bibitem{Dir38} P.A.M. Dirac, Proc. R. Soc. London \textbf{A167}, 148 (1938).
\bibitem{QuiWal97} T.C. Quinn and R.M. Wald, Phys. Rev. \textbf{D56}, 3381
(1997).
\bibitem{BurLiuSoe00} L.M. Burko, Y.T. Liu and Y. Soen, Phys. Rev. \textbf{D63},
124015 (2000).
\bibitem{BurLiu01} L.M. Burko, Y.T. Liu, Phys. Rev. \textbf{D64},
024006 (2001).
\bibitem{Khu03} N.R. Khusnutdinov, Phys. Rev. \textbf{D67}, 124020 (2003); J.
Math. Phys. \textbf{44}, 2320 (2003).
\bibitem{Vil79} A. Vilenkin, Phys. Rev. \textbf{D20}, 373 (1979).
\bibitem{Lin76} B. Linet, J. Phys. A: Math. Gen. \textbf{9}, 1081 (1976).
\bibitem{BatErdV1} H. Bateman, A. Erdelyi, Higher Transcedental
Functions, V.1, (Mc Graw-Hill Book Company Inc, 1953).
\bibitem{Vass03} D. Vassilevich, Phys. Rep. \textbf{388}, 279 (2003).
\end{thebibliography}
\end{document}